\documentclass{PoS}

\title{Diseases with rooted staggered quarks}

\ShortTitle{Diseases with rooted staggered quarks}

\author{\speaker{Michael Creutz}\\ %\thanks{A footnote may follow.}\\
        Brookhaven National Laboratory, Upton, NY 11973, USA\\
        E-mail: \email{creutz@bnl.gov}}

\abstract{Calculations using staggered quarks augmented with a root of
 the fermion determinant to reduce doubling give a qualitatively
 incorrect behavior in the small quark mass region.  Attempts to
 circumvent this problem for the continuum limit involve an unproven
 combination of unphysical states, a loss of unitarity, and a rather
 peculiar non-commutation of limits.}

\FullConference{XXIV International Symposium on Lattice Field Theory\\
		 July 23-28 2006\\
		 Tucson Arizona, US}

\begin{document}

A popular approximation \cite{Aubin:2004ej,Aubin:2004fs} in lattice
gauge theory arises from the simplicity of the staggered fermion
formulation \cite{ Kogut:1974ag,Susskind:1976jm,Sharatchandra:1981si}.
With only one Dirac component on each site, the large matrix
inversions involved with conventional algorithms are substantially
faster than with other approaches.  However the approach and its
generalizations are based on a discretization method that inherently
requires a multiple of four fundamental fermions.  The reasons for
this are related to the cancellation of chiral anomalies.  To apply
the technique to the physical situation of two light and one
intermediate mass quark requires an extrapolation down in the number
of fermions.

As usually implemented, the approach involves taking a root of the
fermion determinant inside standard hybrid Monte Carlo simulation
algorithms.  This step has not been justified theoretically.  Here I
argue that at finite lattice spacing this gives a qualitatively
incorrect behavior as a function of the quark mass.  To produce a
correct continuum limit requires the removal of unphysical
singularities which persist at finite volume and in regions of
parameter space where there are no physical massless particles.  Even
in regions where the fermion determinant is positive definite, the
procedure imposes non-trivial Ward identities incorporating unphysical
states.
%\footnote{ Doubt is not a pleasant condition, but certainty is absurd.
% --   Voltaire (1694 - 1778)}

\begin{figure*}
\centering
\includegraphics[width=2.5in]{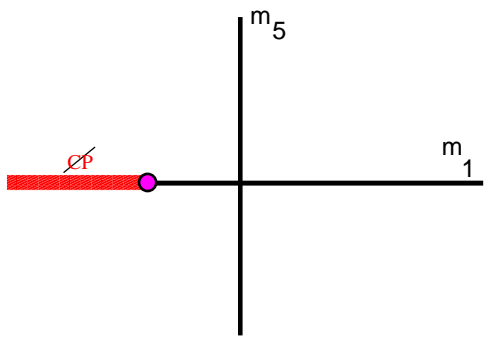}
\caption{ The phase diagram for one flavor quark-gluon dynamics as a
function of $m_1$ and $m_5$ defined in the text.  A first order phase
transition line ending at a second order phase transition appears
along the negative mass axis.  The transition represents a spontaneous
breaking of CP symmetry.  A finite gap separates the transition from
vanishing quark mass.}
\label{oneflavor} 
\end{figure*}

The basic problem arises from the exact chiral symmetry of staggered
quarks.  This symmetry appears for each flavor independently and
survives the rooting procedure.  Due to well known anomalies, this
symmetry can exceed that allowed in the target theory.  

The issues are sharpest in one flavor QCD.  This theory has a rather
interesting phase diagram when considered as a function of the quark
mass \cite{DiVecchia:1980ve,Leutwyler:1992yt,Creutz:2000bs}.  In
particular, consider the generalized mass term in continuum notation
\begin{equation}
 m_1\ \ \overline\psi\psi
 +i\ m_5\ \ \overline\psi\gamma_5\psi
\label{mfive}
\end{equation}
As a function of $m_1$ and $m_5$, this model has a first order phase
transition along a portion of the negative $m_1$ axis, as sketched in
Fig.(\ref{oneflavor}).  In many discussions $m_1$ and $m_5$ are
combined to define a ``complex'' mass $m=m_1+im_5$.

The $m_5$ term introduces an explicit breaking of $CP$ symmetry.  When
$m_5$ vanishes, the first order transition along the negative mass
axis represents a spontaneous breakdown of this symmetry.  A
convenient order parameter is the pseudoscalar expectation
\begin{equation}
\langle\eta^\prime\rangle=i\langle \overline\psi\gamma_5\psi\rangle.
\end{equation}
This quantity displays a finite jump in sign as one passes through the
transition.

This phase diagram illustrates several important features of one
flavor QCD.  First, the naive classical symmetry under $m\rightarrow
-m$ is lost, having been broken by the anomaly.  Second, this
transition line does not start at $m=0$ but is separated from that
point by a finite gap.  Indeed, physics is expected to be analytic in
the mass for a finite region around the origin.  The main point I am
raising is that these features are missing with rooted staggered
quarks at finite lattice spacing.  For the correct physics to be
recovered in the continuum would require a complex and implausible
cancellation of unphysical effects.
%\footnote{
%There are few nudities so objectionable as the naked truth.
% --   Agnes Repplier (1855 - 1950)
%}

I briefly review the rooted staggered quark approach.  Starting with
naive lattice fermions, well known doubling issues give the fermion
spectrum a sixteen fold degeneracy.  The staggered approach reduces
this degeneracy using the observation that for any closed fermion
loop, the product of gamma matrices associated with that loop is
proportional to the identity.  If a fermion loop starts with only one
non-vanishing component on any site, it always involves only a single
component on each site it passes through.  The sixteen doublers
immediately factor into four independent pieces.  This can be
formalized by introducing a projection operator on each site
\begin{eqnarray}
\psi\rightarrow P\psi\\
P=P^2\\
{\rm Tr}\ P=1
\end{eqnarray}
A particular implementation of this is
\begin{equation}
P={1\over 4} \bigg(1+i\gamma_1\gamma_2 (-1)^{x_1+x_2}
+i\gamma_3\gamma_4 (-1)^{x_3+x_4}
 +\gamma_5(-1)^{x_1+x_2+x_3+x_4}\bigg)
\end{equation}
where the $x_i$ are the integer coordinates of the lattice site in
question.  This projection reduces the degeneracy from sixteen to
four.

To reduce the doubling further requires something additional, and this
is where the rooting ``trick'' comes in.  Formally the procedure
starts with the path integral and replaces the fermion determinant
with its fourth root, i.e. $|D|\rightarrow |D|^{1/4}$.  In practice
this is achieved by inserting a factor of $1/4$ in the fermion force
term while traversing a hybrid Monte Carlo trajectory.  As $D$ still
only involves a single fermion component on each site, the process
remains quite frugal with computer time.  Indeed, this is the main
motivation for the scheme.

The problems with this stem from an exact symmetry of the unrooted
theory that should not be there for the one flavor model.  To see this
symmetry explicitly in terms of the basic parameters, generalize the
staggered mass term to
\begin{equation}
( m_1 + i S(x)\ m_2)\ \overline\psi(x)\psi(x)
\label{mtwo}
\end{equation}
where $S(x)$ represents the parity of the site in question,
\begin{equation}
S(x) = (-1)^{x_1+x_2+x_3+x_4}.
\end{equation}
With this extension, the fermion determinant and therefore the path
integral are exactly invariant under the transformation
\begin{equation}
\pmatrix{m_1\cr m_2\cr}
\longrightarrow
\pmatrix{\cos(\theta) & \sin(\theta)\cr
-\sin(\theta) & \cos(\theta)\cr}
\pmatrix{m_1\cr m_2\cr}
\label{rotation}
\end{equation}
As the mass becomes small, the symmetry requires the appearance of a
Goldstone boson or parity doubling.  For the unrooted theory this is
not a problem.  The various doublers rotate differently under chiral
symmetry, and this represents a non-singlet axial symmetry.  The
Goldstone boson is one of the expected pseudoscalar bound states of
the four fermion ``tastes.''
%\footnote{
%Actions lie louder than words.
%--    Carolyn Wells
%}

Is there some connection between the parameter $m_5$ in
Eq.~(\ref{mfive}) and the parameter $m_2$ in Eq.~(\ref{mtwo})?
Naively there would seem to be since the projection operator satisfies
\begin{equation}
\gamma_5 P = S(x) P
\end{equation}
However for the unrooted theory the rotation in Eq.~(\ref{rotation})
is a flavored axial symmetry; so, there is a hidden flavor matrix in
Eq.~(\ref{mtwo}).  For the rooted theory this symmetry is spurious,
the theories are different, and the question of this paragraph is
moot.

To make the problem with rooting more precise, consider working in
finite volume and on a lattice with an even number of sites.  In that
case the determinant of the fermion matrix $|D|$ is a polynomial in
the combination $m_1^2+m_2^2$.  This is easily seen from the hopping
parameter expansion.  Furthermore $|D|$ is real and non-vanishing for
real $m_1^2+m_2^2>0$.  This comes by looking at the eigenvalues of $D$
after rotating away $m_2$ in favor of $m_1$.  Finally, there is a
lower bound on the determinant
\begin{equation}
|D|\ge (m_1^2+m_2^2)^{N_c V/2}
\end{equation} 
where $N_c$ is the number of colors and $V$ the number of lattice
sites.

These facts show that roots of the determinant are analytic in $m_1$
and $m_2$ for at least some finite vicinity of any point where
$m_1^2+m_2^2>0$.  This analyticity allows a continuation from $m$ to
$-m$.  For example one can follow the path
\begin{equation}
m_1=m\cos(\theta),\qquad m_2=m\sin(\theta),\qquad 0\le\theta\le\pi.
\end{equation}
This continuation avoids all singularities in $|D|^{1/4}$.

We conclude that for rooted staggered fermions all correlations at
$m_1$ are identical to those at $-m_1$.  Furthermore the Ward
identities that follow from ${d^n\over d^n\theta} Z =0$ require a
Goldstone boson or parity doubling.  These properties are in direct
contradiction to the known behavior of the one flavor theory, which is
analytic at $m=0$ with the $m$ and $-m$ theories being inequivalent
and there is no Goldstone boson or parity doubling.  While the one
flavor theory is somewhat peculiar, these problems extend directly to
more flavors with non-degenerate quarks.  This is because one has an
independent $U(1)$ symmetry for each flavor, which is one more
symmetry than allowed.

The advocates for staggered quarks \cite{Bernard:2006vv} have argued
that these facts would be irrelevant if the diseases were to go away
in the limit of vanishing lattice spacing.  They argue that the the
limits $a\rightarrow0$ and $m\rightarrow 0$ might not commute.  They
point out that instanton effects can bring in $\sqrt{m^2+a^2}$ factors
which break the $m\rightarrow -m$ symmetry if $a\rightarrow 0$ first.
They do allow that there are extra Ward identities, but these could
be satisfied by indefinite metric ``taste'' states.  The required
Goldstone boson would be there, but as one of a larger set of
unphysical states.  The use of an indefinite metric could permit a
cancellation of these extra states from all physical processes.
%\footnote{
%The important thing is not to stop questioning.
% --   Albert Einstein
%}

Is this picture for saving the rooting procedure plausible?  The idea
of infrared and ultraviolet limits not commuting seems a bit strange,
but the staggered proponents are asking a lot more.  The problems
persist even at finite volume, where one would not expect any infrared
issues with the mass going to zero.  Furthermore, for the one flavor
theory the physical spectrum does not contain any massless particles;
so, it is unclear why an infrared regulator should be required anyway.

Note that the chiral behavior at finite cutoff $a$ is explicitly wrong
because of the extra Goldstone boson; indeed, this invalidates what
was once regarded as a virtue of staggered fermions, the fact that
they have some remnant chiral symmetry.  Furthermore, the use of an
indefinite metric for unphysical states requires
a major leap from the canonical lattice ideas of Osterwalder
and Seiler \cite{Osterwalder:1977pc}.  Finally one should note that
these issues are not present with other fermion regulators, such as
Wilson \cite{Wilson:1975id}, domain wall \cite{Furman:1994ky}, 
or the overlap \cite{Neuberger:1997fp}.

In conclusion, at finite lattice spacing rooted staggered quarks give
a qualitatively incorrect description of one flavor QCD.  Attempts to
circumvent this require complicated and implausible limits, not
necessary with other lattice regulators.  Without proof, rooted
staggered quarks should be regarded as an uncontrolled approximation,
from which claims of first principles calculations are at best
premature.
%\footnote{
%Love truth, and pardon error.
%--    Voltaire (1694 - 1778)
%}

\section*{Acknowledgments}
This manuscript has been authored under contract number
DE-AC02-98CH10886 with the U.S.~Department of Energy.  Accordingly,
the U.S. Government retains a non-exclusive, royalty-free license to
publish or reproduce the published form of this contribution, or allow
others to do so, for U.S.~Government purposes.

\end{document}